\documentclass[conference]{IEEEtran}
\IEEEoverridecommandlockouts
\usepackage[ruled,vlined]{algorithm2e}
\usepackage{subcaption}
\usepackage{bbm}
\usepackage{adjustbox}
\usepackage{lipsum}
\usepackage{stmaryrd}
\usepackage{bm}
\usepackage{cite}
\usepackage{makecell}   
\usepackage{amssymb,amsfonts}
\usepackage{pifont}
\usepackage{tabularx}
\newcolumntype{Y}{>{\centering\arraybackslash}X}
\usepackage{booktabs} 
\usepackage{algorithmic}
\usepackage{graphicx}
\usepackage{siunitx}
\usepackage{amsmath}    
\usepackage{hyperref}
\usepackage{textcomp}
\usepackage{svg}
\usepackage{float}
\usepackage{xcolor}
\def\BibTeX{{\rm B\kern-.05em{\sc i\kern-.025em b}\kern-.08em
    T\kern-.1667em\lower.7ex\hbox{E}\kern-.125emX}}
\begin{document}
\title{\huge Multi-Agent Reinforcement Learning for Joint Handover Management and Power Allocation\\ in Multi-Orbit Satellite Networks }

\author{Yassine Afif$^{*\ddag}$, Ashutosh Balakrishnan$^@$, Philippe Martins$^@$, Mohammed Almekhlafi$^*$,\\
Antoine Lesage-Landry$^{*\ddag}$ and Gunes Karabulut Kurt$^*$\\
$^*$Department of Electrical Engineering, Polytechnique Montréal \& Poly-Grames Research Centre, QC, Canada\\
$^\ddag$Mila \& GERAD, Montréal, QC, Canada\\
$^@$Department of Computer Science and Networks, IMT-Télécom Paris, Palaiseau, France\\
Emails: \{yassine.afif, mohammed.al-mekhlafi, antoine.lesage-landry, gunes.kurt\}@polymtl.ca\\
\{balakrishnan, martins\}@telecom-paris.fr
}

\maketitle
\IEEEpeerreviewmaketitle
\begin{abstract}

Future sixth-generation non-terrestrial networks are expected to combine low Earth orbit (LEO), medium Earth orbit (MEO), and geostationary Earth orbit (GEO) satellites, whose complementary layers must be coordinated through joint user association, power allocation, and handover management under fast LEO dynamics. This paper studies this problem by formulating it as a mixed-integer nonlinear program and decomposing it into a multi-agent reinforcement learning (MARL) policy that selects the associations and a convex power-allocation subproblem solved exactly at each time slot that defines the reward of the MARL part. The association policy is trained with multi-agent proximal policy optimization (MAPPO) and the targeted multi-agent communication (TarMAC) mechanism, and is made aware of the orbital layer through a state that encodes layer-dependent handover penalties. Evaluated on a realistic multi-constellation scenario built from real two-line element data over Nairobi, Kenya, the proposed policy reaches $92\%$ of the throughput of a greedy signal-to-noise ratio (SNR) maximizing scheme while triggering more than four times fewer handovers, and improves throughput by roughly $14\%$ over a conservative stay heuristic. Compared to an LEO-only learned policy of identical architecture, it attains slightly higher throughput with fewer handovers by offloading a fraction of the users to the MEO and GEO layers, an emergent multi-orbit behavior that drives its favorable throughput and handover trade-off.
\end{abstract}


\section{Introduction}

The vision of sixth-generation (6G) wireless systems pushes connectivity beyond terrestrial infrastructure toward ubiquitous, three-dimensional coverage for remote, oceanic, and aerial regions where ground deployment is economically or physically infeasible. Non-terrestrial networks (NTNs) have consequently evolved from a complementary technology into a core component of future wireless networks~\cite{6G,survey,11021288}. Among the various NTN platforms, satellite constellation-based networks have emerged as a key enabler of global connectivity, motivated by recent advances in satellite manufacturing, launch capabilities, and network integration~\cite{6G}.

Satellite systems are usually classified based on their orbital altitude into three main categories: low Earth orbit (LEO), medium Earth orbit (MEO), and geostationary Earth orbit (GEO). LEO satellites operate below $2{,}000$ km, providing low propagation delays and high data rates but at the cost of limited coverage and rapid relative motion \cite{survey}. MEO satellites are placed at intermediate altitudes between $2{,}000$ km and $35{,}786$ km, balancing between latency and coverage. GEO satellites, located at an altitude of approximately $35{,}786$ km, enable persistent wide-area coverage but at the expense of higher propagation delays \cite{11021288}. These orbital regimes exhibit distinct characteristics in terms of coverage, latency, visibility duration, and resource availability, making them naturally complementary components of future NTN architectures~\cite{maral2020satellite}.

Motivated by these diverse characteristics, recent research increasingly considers them as a unified multi-layer satellite infrastructure. The different orbital regimes involve complementary trade-offs that can be jointly exploited to improve network performance. Large LEO constellations such as Starlink and Kuiper can provide the primary source of high-capacity connectivity, while MEO and GEO satellites can support traffic demand through dynamic traffic offloading mechanisms~\cite{offloading}. Such cooperation allows higher orbital layers to compensate for temporary LEO congestion, coverage gaps, or service disruptions caused by the rapid movement of LEO satellites. 

Accordingly, this integration of heterogeneous satellite layers offers significant opportunities to improve coverage, capacity, and service continuity in future NTNs. Realizing these benefits requires efficient resource management mechanisms capable of coordinating network operations across multiple layers with distinct characteristics and constraints. Factors such as satellite mobility, heterogeneous coverage footprints, limited onboard resources, and dynamic traffic demands introduce complex interactions that must be jointly considered. Consequently, advanced optimization and resource allocation strategies are becoming essential for fully leveraging the potential of integrated multi-layer satellite networks~\cite{survey}. In particular, power allocation and handover management have emerged as tightly coupled optimization problems whose impact becomes increasingly significant as the scale and heterogeneity of satellite constellations continue to grow.

\subsection{Related Work}
\label{sec:related-work}
Resource-efficient connectivity through LEO mega-constellations raises three coupled challenges: handling frequent handovers, exploiting multiple orbital layers, and coordinating distributed resource-allocation decisions at scale. We review these three areas in turn before positioning our contributions.

\subsubsection{Handover Management in LEO Constellations}

The high mobility of LEO satellites forces frequent handovers and expensive signaling overheads. Heuristic and game-theoretic methods address this through load balancing~\cite{wu2019potential} or comparative analysis across LEO, MEO, and  highly elliptical orbit constellations~\cite{voicu2024handover}. Optimization-based methods formulate the handover problem as a network flow~\cite{zhang2021networkflows} or as a mixed-integer problem, coupling association and power allocation with a handover penalty~\cite{afif2025}. Solving each slot independently, however, neglects the sequential nature of the problem, which reinforcement-learning (RL) approaches are better suited to capture. Our earlier single-layer study~\cite{afif2026marlleo} follows this direction, casting joint beam association and power allocation in LEO networks as a multi-agent RL (MARL) problem with explicit and learned attention-based inter-agent communication. The study in~\cite{lee2024dho} proposes a handover protocol that skips the measurement report phase to reduce access delay and collisions. The authors of~\cite{tong2025a2c} adopt the advantage actor-critic algorithm to scale to very large-scale constellations. The framework proposed in~\cite{he2020loadaware} distributes the decision across agents via MARL, foreshadowing the methods discussed below.

\subsubsection{Multi-Orbit Integration}
While the works above treat the LEO layer in isolation, additional flexibility arises once several orbits are considered jointly within NTNs. Reference~\cite{azari2022evolution} surveys the evolution of NTNs from fifth-generation to 6G, and the study in~\cite{lin2021path} positions LEO satellite access as a radio-access building block. Exploiting multi-orbit complementarity (LEO+MEO+GEO), the approach in~\cite{jiang2020capacity} proposes RL-based capacity management in a multi-layer network, while the authors in~\cite{han2025multiorbit} design a multi-orbit soft-handover strategy based on rate-splitting multiple access. These works show that resource allocation should be considered jointly across orbits rather than within an isolated layer.

\subsubsection{MARL for Resource Management in Networks}
The multi-orbit network perspective calls for a decentralized coordination among many agents, the setting that MARL addresses. RL is commonly extended to the MARL setting via the centralized-training, decentralized-execution paradigm~\cite{kraemer2016multiagent}, e.g., multi-agent deep deterministic policy gradient~\cite{lowe2017maddpg} and multi-agent proximal policy optimization (MAPPO) \cite{yu2022mappo}. This extension however remains agnostic to the interaction structure between agents. Conversely, MARL can also endow agents with communication and structural awareness. For example, differentiable inter-agent learning~\cite{foerster2016dial} and CommNet~\cite{sukhbaatar2016commnet} learn what to communicate, and targeted multi-agent communication (TarMAC)~\cite{das2019tarmac} learns message content and recipients jointly. When the inter-agent interaction structure is represented as a graph, with each agent as a node and its neighbours as edges, graph convolutional reinforcement learning (DGN)~\cite{jiang2020dgn} applies graph convolutions over each agent's dynamically changing neighbourhood to extract cooperative behaviour. These principles have been transposed to satellite networks: \cite{hu2020payload} controls a flexible payload with deep MARL; \cite{lin2022beamhopping} allocates beam pattern and bandwidth with multiple agents; and~\cite{he2022attention} combines a multi-agent actor-critic with attention for satellite-terrestrial resource allocation, bridging generic MARL advances and the constraints of satellite constellations.

 Prior work typically addresses these aspects in isolation: optimization-based methods cover handover and power allocation but use neither learning nor multiple agents, learned handover schemes rarely span multiple orbits or use inter-agent communication, and MARL-based resource-allocation methods rarely integrate handover in a multi-orbit setting. To the best of our knowledge, no prior work has jointly considered handover-aware, multi-orbit resources and power allocation using a communication-enabled MARL framework.

\subsection{Contributions}
The contributions of this paper are as follows.
\begin{itemize}
  \item We formulate the joint satellite association, power allocation, and
        handover management problem for multi-layer (LEO/MEO/GEO)
        constellations as a hierarchical decomposition problem: first, a combinatorial association
        decision solved by a MARL policy, followed by a convex power-allocation
        problem solved in closed loop. The
        handover penalties are present in both the convex objective and
        the learning reward.
  \item We build an evaluation pipeline on a realistic multi-constellation
        scenario, built from real data for a ground deployment in
        Nairobi, Kenya, that goes beyond standard throughput and handover metrics and
        explicitly measures how the learned policy distributes user
        associations across the LEO, MEO, and GEO layers, revealing an
        emergent offloading behaviour.
\end{itemize}
The remainder of the paper is organized as follows. Section~\ref{sec:model} introduces the system model and the joint optimization problem. Section~\ref{sec:approach} describes the proposed layer-aware MARL approach. Section~\ref{sec:setup} details the experimental setup, introduces the baselines and discusses the results. Finally, Section~\ref{sec:conclusion} concludes the paper.
\section{System Model and Problem Formulation}
\label{sec:model}
This section describes the multi-layer satellite network considered in this work and formulates the joint association, power-allocation, and handover problem.
\subsection{Network Topology}
\begin{figure}[t]
  \centering
  \includegraphics[width=.9\linewidth]{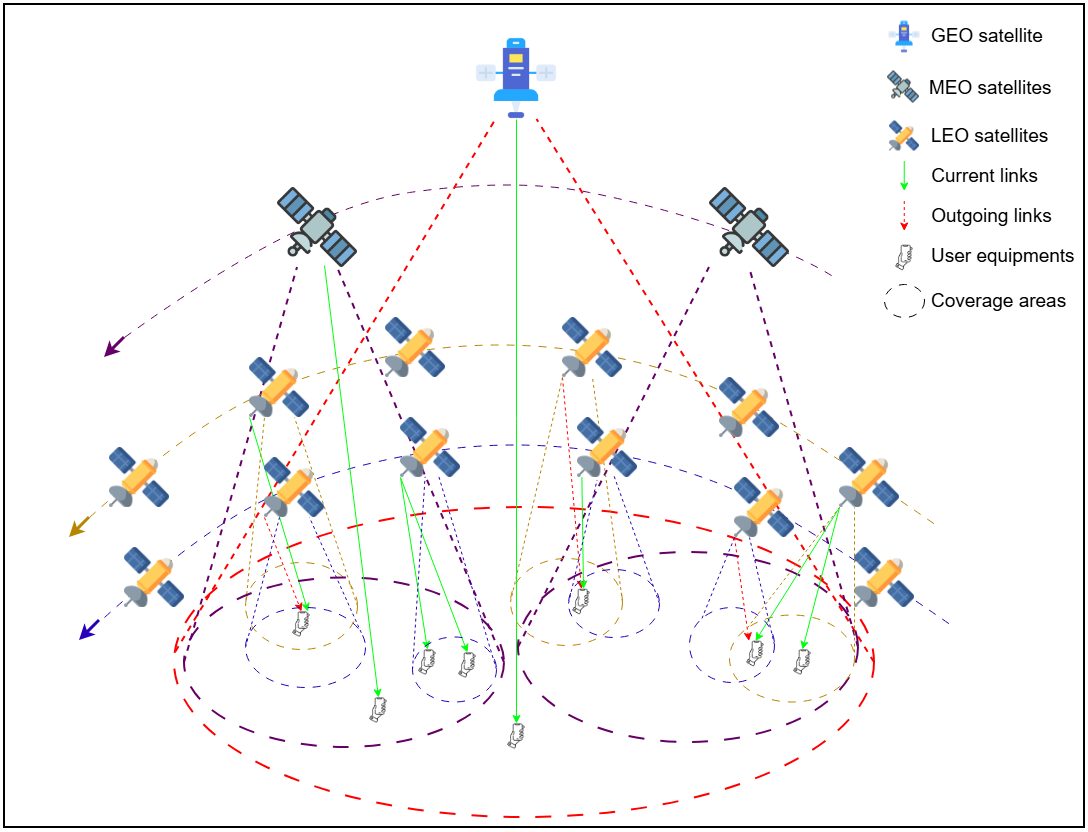}
  \caption{Multi-layer satellite network: $U$ ground users are served by
  satellites from the LEO, MEO, and GEO layers.}
  \label{fig:system-model}
  \vspace{-.25 in}
\end{figure}

We consider a multi-layer satellite network serving a set $\mathcal{U} = \{1, \dots, U\}$ of $U\in\mathbb{N}$ single-antenna ground users, as shown in Fig.~\ref{fig:system-model}. The constellation comprises a set $\mathcal{S} = \{1, \dots, S\}$ of $S\in\mathbb{N}$ satellites, partitioned into three orbital layers $\ell \in \mathcal{L} = \{\text{LEO}, \text{MEO}, \text{GEO}\}$. The superscript $\ell$ on a parameter means it depends on the orbital layer of the satellite or of the user's serving satellite. Each satellite is equipped with $B\in\mathbb{N}$ beams indexed by $\mathcal{B} = \{1, \dots, B\}$. The  time horizon is divided into discrete slots $t \in \mathcal{T} = \{1, \dots, T\}$, where $T\in\mathbb{N}$. To model satellite-user association, we define a binary association variable $I_{u,s,b}^{(t)} \in \{0,1\}$ as
\begin{equation}\label{eq:assoc}
I_{u,s,b}^{(t)} =
\begin{cases}
1, & \text{if }u \text{ is linked to } (s,b) \text{ at } t, \\
0, & \text{otherwise}.
\end{cases}
\end{equation}
At each discrete time slot $t$, the geometry (elevation angle, slant range) between every user-satellite pair is computed from orbital propagation. This yields a time-varying visibility indicator $V_{u,s}^{(t)} \in \{0,1\}$, defined as
\begin{equation}\label{eq:visibility}
V_{u,s}^{(t)} =
\begin{cases}
1, & \text{if } \theta_{u,s}^{(t)} \geq \theta_{\text{vis}}, \\
0, & \text{otherwise},
\end{cases}
\end{equation}
where $\theta_{u,s}^{(t)}$ denotes the elevation angle of pair $(s, u)$ at time $t$, and $\theta_{\text{vis}}$ is a layer-dependent visibility threshold.
\subsection{Rate Analysis}
\label{subsec:channel}
 The channel model, considered in this work, captures both large-scale propagation effects and small-scale fading~\cite{goldsmith2005wireless,itu618,3gpp-38811,iturpy-2017}. The small-scale fading is modeled as Rician fading with a sufficiently large Rician factor $K$, representing the dominant line-of-sight (LOS) component characteristic of the quasi-optical propagation conditions encountered in satellite communications~\cite{itu618,3gpp-38811,iturpy-2017}, and a weaker non-LOS (NLOS) multipath term representing the scattering effects. The large-scale propagation effects include free-space path loss, atmospheric attenuation, ionospheric and tropospheric scintillation, and rain attenuation. Let $|h_{u,s,b}^{(t)}|^2$ and $P_{u,s,b}^{(t)}$ represent the channel gain and the transmit power on link $(u,s,b)$ for all $u \in \mathcal{U}$, $s \in \mathcal{S}$ and $b \in \mathcal{B}_s$ at time slot $t$. Then, the received signal-to-noise ratio (SNR) is
\begin{equation}\label{eq:gamma}
    \gamma_{u,s,b}^{(t)} = \frac{G_{\text{tx}}^{(\ell)}\, |h_{u,s,b}^{(t)}|^2\, P_{u,s,b}^{(t)}}{N_0\, W\, L_{\text{total,lin}}^{(u,s,t)}},
\end{equation}
where $N_0$, $W$, and $L_{\text{total,lin}}^{(u,s,t)}$ are the noise spectral density, bandwidth, and the total path loss, respectively.  $G_\text{tx}^{(\ell)}$ denote the antenna gain of layer $\ell\in\{\text{LEO}, \text{MEO}, \text{GEO}\}$, which is such that :  
\begin{equation}\label{eq:gains}
    G_{\text{tx}}^{\text{LEO}} < G_{\text{tx}}^{\text{MEO}} < G_{\text{tx}}^{\text{GEO}},
\end{equation}
where transmit antenna gains are assumed to increase with orbital altitude in order to compensate for the greater propagation losses experienced by higher-layer satellites. Accordingly, the achievable Shannon rate of link $(u,s,b)$ in layer $\ell$ at time slot $t$, for all $u \in \mathcal{U}$, $s \in \mathcal{S}$, and $b \in \mathcal{B}_s$, is given by
\begin{equation}\label{eq:rate_link}
    R_{u,s,b}^{(t)} = W\, \log_2\!\left(1 + \gamma_{u,s,b}^{(t)}\right).
\end{equation}
Accordingly, the achievable rate of user $u$ aggregates the contributions of its associated links only, and can be written as
\begin{equation}\label{eq:rate}
    R_u^{(t)} = \sum_{s\in\mathcal{S}} \sum_{b\in\mathcal{B}} R_{u,s,b}^{(t)}.
\end{equation}
\subsection{Handover Modelling}
\label{subsec:handover}
In a multi-layer constellation, the LEO layer in particular exhibits a high orbital velocity, causing its visible-satellite set to change rapidly and forcing frequent reassociations, whereas MEO and GEO satellites remain visible over longer periods. A handover occurs when a user $u\in\mathcal{U}$ switches its serving beam between two consecutive slots. We define the handover indicator at slot $t \in \{2, \dots, T\}$ as
\begin{equation}\label{eq:handover}
H_u^{(t)} = 1 - \sum_{s\in\mathcal{S}} \sum_{b\in\mathcal{B}} I_{u,s,b}^{(t)}\, I_{u,s,b}^{(t-1)},
\end{equation}
so that $H_u^{(t)} = 0$ when the user stays on the same beam and $H_u^{(t)} = 1$ otherwise. To distinguish handover types, let $\delta_u^{(t)} = |\ell_u^{(t)} - \ell_u^{(t-1)}|$ be the corresponding layer gap. The handover penalty is then
\begin{equation}\label{eq:beta}
\beta_u^{(t)} = H_u^{(t)}
\begin{cases}
\alpha_1, & \text{if } \delta_u^{(t)} = 0 \quad \text{(intra-layer)} \\
\alpha_2, & \text{if } \delta_u^{(t)} = 1 \quad \text{(adjacent-layer)} \\
\alpha_3, & \text{if } \delta_u^{(t)} = 2 \quad \text{(LEO } \text{to } \text{GEO)},
\end{cases}
\end{equation}
with $0 \leq \alpha_1 < \alpha_2 < \alpha_3 < 1$, so that larger altitude gaps incur longer service interruptions.
\subsection{Problem Formulation}
\label{pb}
At each time slot $t$, the association variables $\mathbf{I}=\{I_{u,s,b}^{(t)}\}$ and power allocation variables $\mathbf{P}=\{P_{u,s,b}^{(t)}\}$ are jointly optimized to maximize the
network throughput while accounting for handover penalties through the layer-dependent factor $\beta_u^{(t)}$ defined in~\eqref{eq:beta}. The optimization problem is formulated as

\begin{subequations}\label{downlink}
\begin{IEEEeqnarray}{l rCl l}
\IEEEeqnarraymulticol{5}{l}{%
  \max_{\mathbf{P},\mathbf{I}}\quad \sum_{t\in\mathcal{T}}\sum_{u\in\mathcal{U}} R_u^{(t)}\bigl(1-\beta_u^{(t)}\bigr)} \label{P1:obj}\\
\;\,\text{s.t.} & R_u^{(t)} &\geq& R_{\text{th}}, & \forall u,t, \label{P1:cons1}\\
& P_{u,s,b}^{(t)} &\leq& I_{u,s,b}^{(t)}\,P_{\max}^{(\ell)}, & \forall u,s,b,t, \label{P1:cons2}\\
& \sum_{u\in\mathcal{U}}\sum_{b\in\mathcal{B}} P_{u,s,b}^{(t)} &\leq& P_{\max}^{(\ell)}, & \forall s,t, \label{P1:cons3}\\
& \qquad\sum_{u\in\mathcal{U}} I_{u,s,b}^{(t)} &\leq& C, & \forall s,b,t, \label{P1:cons4}\\
& \qquad\sum_{s\in\mathcal{S}}\sum_{b\in\mathcal{B}} I_{u,s,b}^{(t)} &\leq& 1, & \forall u,t, \label{P1:cons5}\\
& I_{u,s,b}^{(t)} &\leq& V_{u,s}^{(t)}, & \forall u,s,b,t, \label{P1:cons6}\\
& P_{u,s,b}^{(t)}&\geq& 0, \, I_{u,s,b}^{(t)} \in \{0,1\}, \,& \forall u,s,b,t, \label{P1:cons7}\\
& \IEEEeqnarraymulticol{4}{l} {\qquad \eqref{eq:visibility},~\eqref{eq:gamma},~\eqref{eq:rate_link},~\eqref{eq:rate},~\eqref{eq:beta}.} \nonumber
\end{IEEEeqnarray}
\end{subequations}
where~\eqref{P1:cons1} enforces that each user's rate satisfies the minimum rate requirement $R_{\text{th}}$. Constraint~\eqref{P1:cons2} guarantees that power is allocated only to active links. Constraint~\eqref{P1:cons3} enforces the per-satellite power budget. Constraint~\eqref{P1:cons4} limits the number of users served by each beam to $C$, while \eqref{P1:cons5} restricts each user to being served by at most one beam. Constraint~\eqref{P1:cons6} allows association only over visible links. Finally, constraint~\eqref{P1:cons7} defines the binary association and non-negative power domains. Problem~\eqref{downlink} is a mixed-integer nonlinear program (MINLP). The non-concavity stems primarily from constraint~\eqref{P1:cons2}, which couples the binary association variables $I_{u,s,b}^{(t)}$ with the continuous powers $P_{u,s,b}^{(t)}$ through a bilinear product, so that the feasible set is non-convex and the joint optimization cannot be solved directly by standard convex methods. To make the problem tractable, we split it into more manageable subproblems, as detailed in the next subsection.

\section{Proposed Framework}
\label{sec:approach}
As discussed in Section~\ref{pb}, the bilinear coupling between the discrete association and the continuous power allocation makes a direct joint solution of~\eqref{downlink} intractable. We therefore decompose~\eqref{downlink} into two sequential subproblems solved in closed loop: a learning-based policy that first selects the association, and a convex program that computes the optimal power for a given association, whose resulting rates feed back into the policy. Unlike the single-layer formulation in~\cite{afif2026marlleo}, where association reduced to selecting a single visible satellite, the multi-layer setting introduces layer-dependent handover penalties $\beta_u^{(t)}$ and a much larger candidate space, motivating this decomposition.

\paragraph{Association via MARL}
The association decisions ${I}_{u,s,b}^{(t)}$ are delegated to a MARL policy with one agent per user and shared weights, trained with MAPPO and the TarMAC mechanism~\cite{das2019tarmac}. Each agent observes a layer-aware state and chooses to stay on its current beam or switch to one of its top-$k$ candidates, thereby handling the non-convex combinatorial part and the handover dynamics driving $\beta_u^{(t)}$. Each agent builds an observation of dimension ${d_{\text{obs}}}=3k+4$: (i) min/max normalized SNRs of its top-$k$ links in $[-1,1]$; (ii) a one-hot stay vector marking the serving beam; (iii) $1/(\Delta t_{\text{ho}}+1)$, the inverse of the elapsed time since the last handover; (iv) the episode fraction spent on the current beam; (v) the achieved rate against the quality-of-service target, $\operatorname{clip}\!\big((R_{u}^{(t)}-R_{\text{th}})/R_{\text{th}},-1,1\big)$; and (vi) a layer encoding ($\text{LEO}=0,\text{MEO}=0.5,\text{GEO}=1$) for the $k$ candidates and the serving satellite, the key addition over the single-layer case. For communication, each agent~$u\in\mathcal{U}$ encodes a hidden state $z_u$ along with a key $\kappa_u$, a query $q_u$, and a message $m_u$. Scaled dot-product attention over the messages of its $n$ nearest geographic neighbours (fixed from user coordinates) produces a context $c_u$, which is concatenated with $z_u$ to output the action, viz., stay or switch among the $k$ candidates, and the critic value.
\begin{figure*}[t]
    \centering

    \begin{subfigure}{0.32\linewidth}
        \centering
        \includegraphics[width=\linewidth]{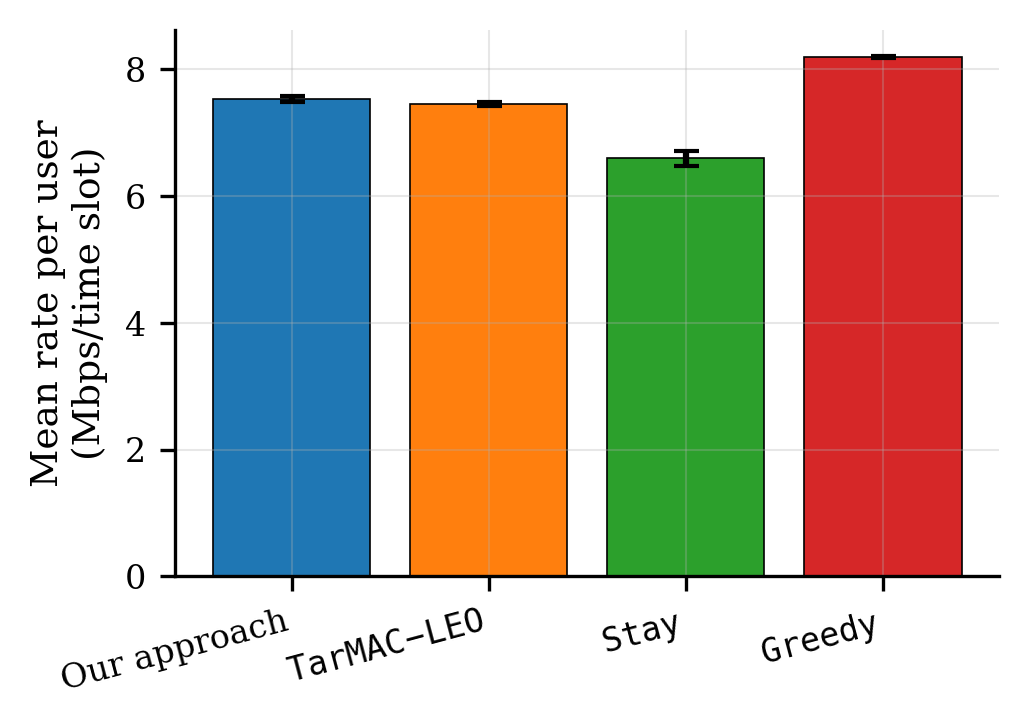}
        \caption{Mean rate per user.}
        \label{fig:throughput_bar}
    \end{subfigure}
    \begin{subfigure}{0.32\linewidth}
        \centering        \includegraphics[width=\linewidth]{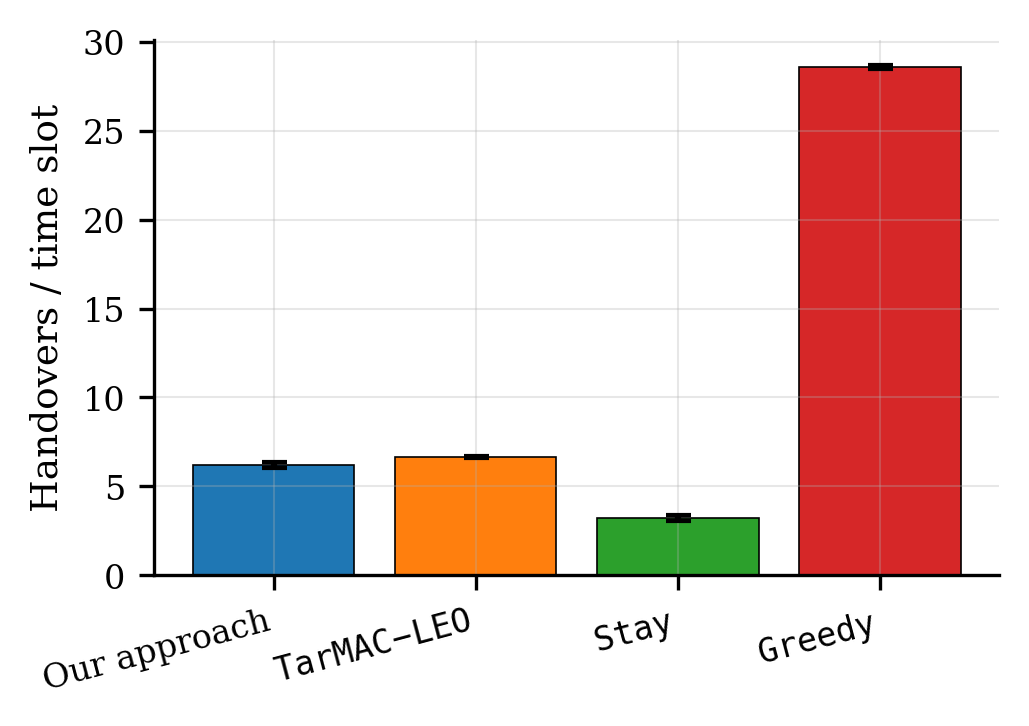}
        \caption{ Number of handovers per time slot.}
        \label{fig:handovers_bar}
    \end{subfigure}
    \begin{subfigure}{0.32\linewidth}
        \centering
        \includegraphics[width=\linewidth]{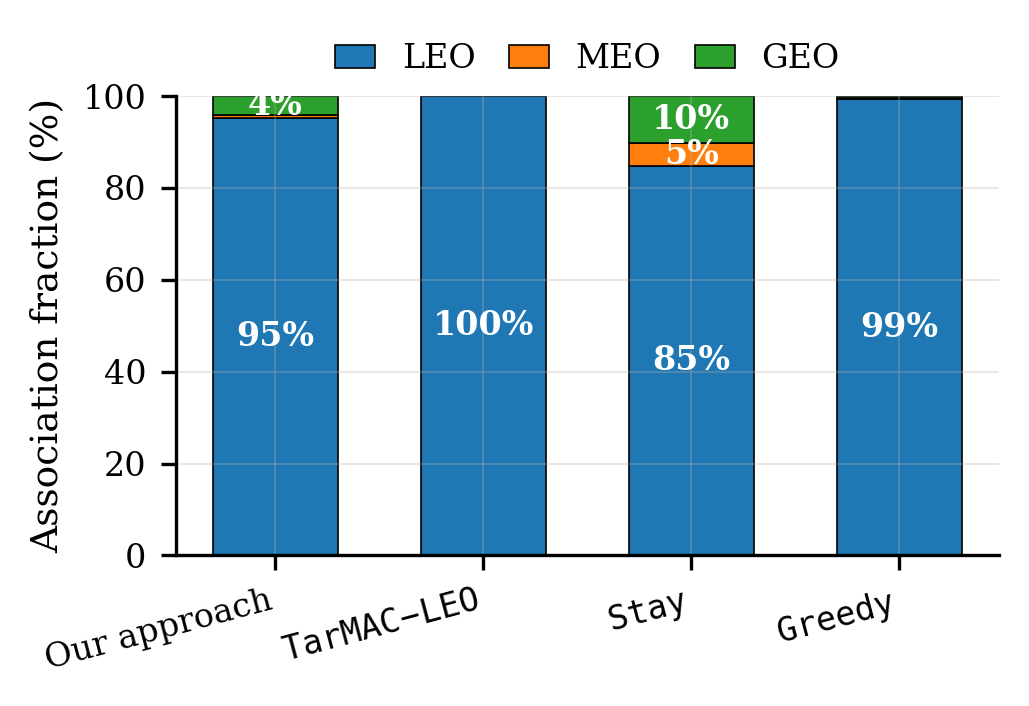}
        \caption{Associations across satellite's layers.}
        \label{fig:layer_utilization}
    \end{subfigure}

    \caption{Performance comparison of the proposed layer-aware policy against the baselines, averaged over $40$ episodes.}
    \label{fig:performance_comparison}
    \vspace{-.25 in}
\end{figure*}
\begin{table}[b]
  \centering
  \caption{ Simulation and training parameters.}
  \label{tab:hyperparams}
  \renewcommand{\arraystretch}{1.3}
  \setlength{\tabcolsep}{4pt}
  \begin{tabular}{ll}
    \toprule
    Parameter & Value \\
    \midrule
    Users $U$ & 30 \\
    Beams $B$ / beam capacity $C$ & 3 / 3 \\
    Episode length $T \times \Delta t$ & 100 $\times$ \SI{30}{\second} \\
    Visibility threshold $\theta_{\text{vis}}$ & \SI{10}{\degree} \\
    Carrier frequency\ $f_{\text{c}}$ / bandwidth $W$ & \SI{20}{\GHz} / \SI{1}{\MHz} \\
    Noise spectral density $N_0$ / Rician factor $K$ & \SI{5e-21}{\watt\per\hertz} / \SI{10}{\dB} \\
    Top-$k$ candidates / neighbors $n$ & 8 / 2 \\
    LEO pool (init / per ckpt / \# ckpts) & 20 / 10 / 40 \\
    GEO / MEO pool size & 5 / 10 \\
    Tx gain $G_{\text{tx}}^{\text{LEO/MEO/GEO}}$ & 32 / 40 / 50 dBi \\
    Power budget $P_{\max}^{\text{LEO/MEO/GEO}}$ & 1k / 15k / 30k W \\
    Minimum rate $R_{\text{th}}$ & 1 Mbps \\
    Handover penalty $\alpha_1/\alpha_2/\alpha_3$ & 0.6 / 0.8 / 0.9 \\
    Training / evaluation episodes & 200 / 40 \\
    \bottomrule
  \end{tabular}
\end{table}
\paragraph{Power allocation}
Once $\mathbf{I}$ is fixed by the MARL algorithm, the resulting problem becomes concave in ${P}_{u,s,b}^{(t)}$, because constraints~\eqref{P1:cons1} to~\eqref{P1:cons3} and~\eqref{P1:cons7} are linear and each $R_u^{(t)}$ is concave through the $\log(1+\gamma_{u,s,b}^{(t)})$ mapping of~\eqref{eq:rate_link}. This subproblem is solved exactly at each time slot with \texttt{CVXPY}~\cite{diamond2016cvxpy} and \texttt{MOSEK}~\cite{mosek}, yielding the optimal power allocation $\mathbf{P}^{\star}$. The resulting rates then define the per-user reward
\begin{equation}\label{eq:reward}
    r_u^{(t)} = R_u^{(t)}\,(1 - \beta_u^{(t)}),
\end{equation}
which is the throughput achieved by user $u$ at slot $t$ from~\eqref{eq:rate}, discounted by the handover penalty $\beta_u^{(t)}$ of~\eqref{eq:beta}, so that the policy is rewarded for the throughput its association decisions enable while being penalized for costly handovers. This closes the loop: the reward, evaluated only after the optimal power allocation, drives the policy update, while the achieved rate relative to the target $R_{\text{th}}$ is encoded in the next-slot state. The policy thus learns associations that yield high throughput once power is optimally allocated.
\section{Numerical Case Study}
\label{sec:setup}
We now assess the proposed framework on a realistic multi-constellation scenario, describing the simulation setup and baselines before presenting the results.
\subsection{Experimental Setup}
We build a multi-constellation scenario from real two-line element (TLE) catalogs~\cite{celestrak} for the LEO, MEO, and GEO layers, with $U=30$ users uniformly deployed within 100 km of Nairobi, Kenya. MEO and GEO pools are fixed sets of satellites selected by average elevation, while the LEO pool is built with a checkpoint based anticipation scheme: starting from $20$ initially visible LEO satellites, $10$ additional satellites are added at each $40$ look-ahead checkpoints over the episode horizon, so the policy is exposed to realistic LEO handover opportunities rather than a static snapshot. Each episode spans $T=100$ decision slots of $\Delta t=\SI{30}{\second}$ and represents a \SI{50}{\minute} period. Table~\ref{tab:hyperparams} summarizes the main simulation and training parameters. The handover-penalty coefficients $\alpha_1, \alpha_2 $ and $\alpha_3$ in~\eqref{eq:beta} grow with the size of the orbital-layer transition: an intra-layer switch ($\alpha_1$) is penalized the least, an adjacent-layer switch ($\alpha_2$) more, and a two-level LEO to GEO switch ($\alpha_3$) the most.

\paragraph{Baselines} We compare our approach against three baselines. Two are standard heuristics: \texttt{stay}, which keeps the current serving satellite as long as it remains visible, and \texttt{greedy}, which always switches to the instantaneously highest SNR visible satellite regardless of handover cost. The third is \texttt{TarMAC-LEO}, a learned single-layer policy carried over from our earlier single-layer work~\cite{afif2026marlleo}, trained with the same architecture and layer-aware features but restricted, throughout training and evaluation, to candidates from the LEO layer only. To ensure a fair comparison, the LEO satellite pool available to this baseline is enlarged with respect to the multi-layer setting: starting from the same $20$ initially visible satellites, $15$ LEO satellites are added at each $40$ look-ahead checkpoints (versus $10$ per checkpoint for the multi-layer policy), so that \texttt{TarMAC-LEO} is not disadvantaged by a sparser candidate set and the comparison isolates the benefit of multi-layer access rather than LEO satellites availability. Comparing the full multi-layer TarMAC against this single-layer variant directly highlights the advantage of giving the policy access to, and awareness of, all three orbital layers simultaneously.

\paragraph{Metrics}  We report four main metrics: the mean per-episode reward (shown as a training curve), the aggregate throughput per time slot and per user, the number of handovers per time slot and per user, and, central to our analysis, the fraction of user associations served by each orbital layer (LEO, MEO, GEO), which quantifies how much each policy relies on the higher layers. We average over the $40$ evaluation episodes with frozen policies (no further learning).

\begin{figure}[tp]
    \centering
    \includegraphics[width=.9\linewidth]{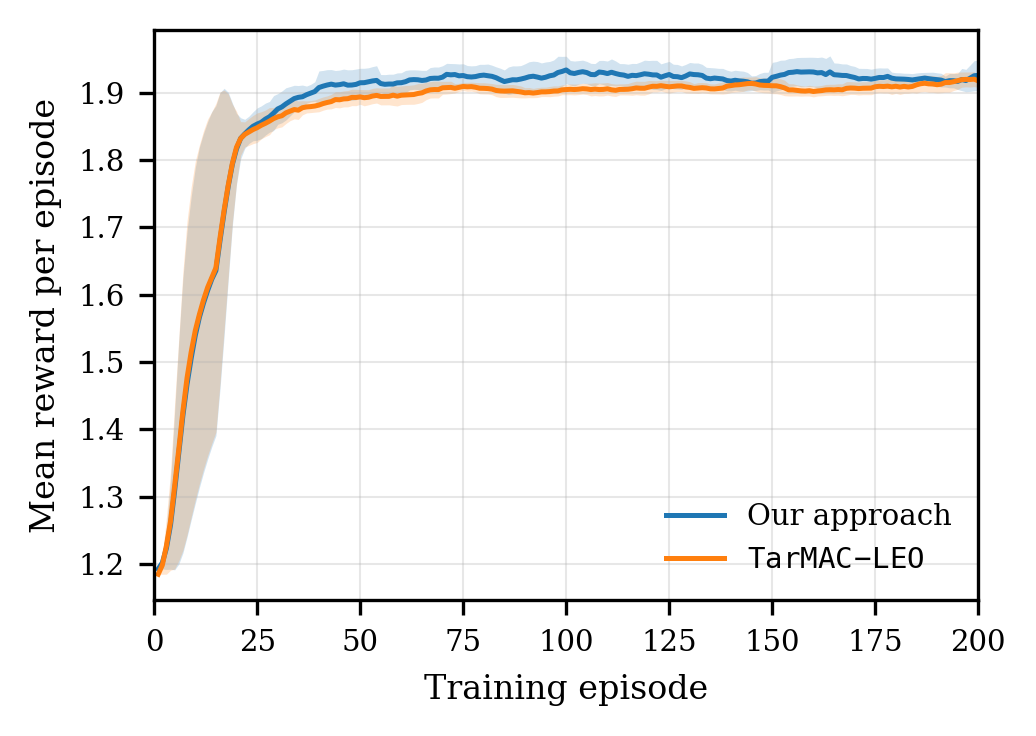}
    \caption{Mean reward per episode during training for the proposed policy and the LEO-only baseline. The shaded area denotes the standard deviation across episodes.}
    \label{fig:reward}
    \vspace{-.25 in}
\end{figure}
\subsection{Results and Discussion}
\label{sec:results}
Fig.~\ref{fig:performance_comparison} compares all policies over $40$ frozen-policy evaluation episodes. The throughput/handover trade-off orders the policies clearly: \texttt{greedy} reaches the highest per-user rate ($8.2$~Mbps per time slot) but triggers $28.5$ handovers per time slot, an order of magnitude more than any learned policy. Our approach sits in the favourable region (Fig.~\ref{fig:handovers_bar}), keeping about $92\%$ of \texttt{greedy}'s rate ($7.55$~Mbps per time slot) while cutting handovers more than four-fold ($6.2$~handovers per time slot), and exceeds the conservative \texttt{stay} heuristic ($6.6$~Mbps per time slot with $3.3$ handovers) by about $14\%$ in rate. Against the LEO-only \texttt{TarMAC-LEO} baseline ($7.48$~Mbps per time slot with $6.7$ handovers), it attains slightly higher rate with fewer handovers. The layer utilization of Fig.~\ref{fig:layer_utilization} reveals why: unlike \texttt{TarMAC-LEO}, locked to $100\%$ LEO, the proposed policy keeps $95\%$ of its associations on LEO and offloads the remaining $5\%$ onto the higher layers ($1\%$ on MEO and $4\%$ on GEO), relieving LEO congestion and outperforming every baseline on the combined rate and handover trade-off.
\section{Conclusion}
\label{sec:conclusion}
This work addressed the joint satellite association, power allocation, and handover management problem in multi-layer low Earth orbit (LEO), medium Earth orbit (MEO), and geostationary Earth orbit (GEO) constellations. The problem is formulated as a mixed-integer nonlinear program. Because the resulting problem is non-convex, we decomposed it into a layer-aware multi-agent reinforcement learning (MARL) association policy and a convex power-allocation subproblem. In a realistic scenario derived from real two-line element data over Nairobi, Kenya, the proposed solution achieved about $92\%$ of the throughput achieved by a greedy policy while reducing more than four-fold handovers, and exceeded a conservative stay heuristic by about $14\%$ in throughput. Compared to an equivalent LEO-only MARL policy, it delivered slightly higher throughput with fewer handovers by offloading part of the traffic onto the MEO and GEO layers, confirming the benefit of exploiting all three orbital layers. Future work will scale the framework to larger user populations in a multi-input multi-output (MIMO) system and incorporate inter-satellite link constraints.

\bibliographystyle{IEEEtran}
\bibliography{refs}

\end{document}